# SAASQUAL: A Quality Model For Evaluating SaaS on The Cloud Computing Environment

[1]Mrs. Dhanamma Jagli, [2]Dr. (Mrs.) Seema Purohit, [3]Dr. N. Subhash Chandra

[1]Research Scholar, JNTU Hyderabad, Telangana, India and
Assistant Professor, V.E.S.Institute of Technology, University of Mumbai, India.
[1]`Email:` dsjagli.vesit@gmail.com

[2] Research Guide, JNTU Hyderabad, Telangana, India and
Kirti College, University of Mumbai, India.
[2]`Email:` supurohit@gmail.com

[3] Research Guide, JNTU Hyderabad, Telangana, India and
[3]Principal and Professor of CSE,Holy Mary Institute of Technology, JNTU Hyderabad, India.
[3]`Email:` subhashchandra_n@yahoo.co.in

**Abstract.** Cloud computing is a Technology that has come out in the last decade and that is transforming the IT industry in huge. The Cloud computing is playing a vital role as a backbone component of the Internet of Things (IoT). In a Cloud Computing scenario, cloud services are accessible via Internet. Cloud computing is providing on-demand resources like Infrastructure, platform, and software as it is do not pay to possess the software itself but rather to use it. Pay for use concept is very attractive, hence many organizations are adopting the SaaS model drastically. Even though, each customer is unique and leads to unique variation in the requirements of the software. the SaaS is generally press into service and it yields advantages to service providers and service customers. More and more SaaS services are emerging, how to select qualified service is key problem for customers. Present quality models are not sufficient to evaluate SaaS selection on the cloud due to its tremendous increasing in the use. A quality model can be used to represent, evaluate, and differentiate the quality of the SaaS providers. In this paper, a new quality model proposed and named SAASQUAL for cloud software services.This model is based on different attributes of quality software, quality service and metrics that measure software quality and service quality in order to evaluate potential software as a service on the cloud.

## I. Introduction

The cloud computing has been growing as an essential and leading computing platform for sharing resources like infrastructure, platform, software, etc. Cloud Computing has emerged as a new paradigm in the field of network-based services within many industries and application domains. The major benefits that it provides in terms of IT efficiency and business agility represent a huge competitive advantage for an organization. This has proven to be an essential requirement for extending many existing applications.The cloud computing is a latest way of computing and providing several scalable resources dynamically as well as virtualizing oftenly those resources as a service over the Internet. There are three main service models are available on the cloud computing environment such as IaaS (Infrastructure as a Service) ,PaaS(Platform as a service) and SaaS(Software as a service).

The SaaS is model provides several uses to service consumers without installing any application locally . In order to use all advantages of SaaS model efficiently and effectively, there should be proper quality model to evaluate SaaS quality.The customers do not pay to possess the software itself but rather to use it,this concept of pay per use is very attractive and commonly used by many users and it had several advantages. In order to make use of the best of SaaS , it is required to examine the possible quality of SaaS. In fact service-providers have to evaluate their services against needs of service-users in order to increase their service. hence the available SaaS quality evaluation models are lacking to focus on all aspects of quality and services together.However, each customer is unique, which leads to a very large variation in the requirements of the software.Therefore, this paper presents a method to help customers to choose a better SaaS product satisfying most of their conditions and alternatives.This is also know that a good method of adaptive selection should be based on the correct definition of the different parameters of choice. For that reason, the proposed work states that extraction and analysis the various parameters involved in the process of the selection of a SaaS Application.

## II. Related Work Towards SaaS Evaluation

The cloud computing will be in need of several numbers of interactions with varying quality requirements. Service quality has become a significant differentiator amid of cloud providers. In order to discriminate between service providers from various competitors, and other cloud service vendors, there should be some measure to know superiority of services.Any standard quality model can be used to represent, measure and compare the quality of the Software services on the cloud. The work has been done many researchers towards SaaS evaluation is comparatively very less till 2005,afterwards work have been improved towards SaaS evaluation till today and many methods were introduced as follows:
Jae Yoo Lee, Jung Woo Lee, Du Wan Chen,"A Quality Model for Evaluating Software-as-a-Service in Cloud Computing". Proposed a quality method that will help to examine the quality aspect of SaaS ,based on the SaaS key features derived from primary SaaS features. They also described some standards to measure quality of SaaS based on the quality attributes. They validated and assessed their proposed quality model for evaluating SaaS' quality,they used the standard EIRE 1061 and examined their work. Manish Godse, Shrikant Mulik, "An Approach for Selecting Software-as-a-Service (SaaS) Product", published and presented a new model based on Analytic Hierarchy Process (AHP) approach intended for ranking the SaaS features like functionality, Vendor Reputation, Architecture, Usability, and Cost.Qian Tao, Huiyou Chang, Yang Yi, Chunqin Gu, "A TRUSTWORTHY MANAGEMENT APPROACH FOR CLOUD SERVICES ",published work,the authors proposed trustworthy management for all cloud services based on non functional QoS attributes like reputation, reliability, security, time, cost and availability. They also applied Data Mining PAM( Partitioning Around Medians) clustering algorithm for trustworthy management of all types cloud services.Chen Yiming, Zhu Yiwei, "SaaS Vendor Selection Basing on Analytic Hierarchy Process ", published work and they proposed a model for selecting Best SaaS vendor based on the same principles applicable to vendors rather than selecting SaaS Product. This model was analyzed by using Analytic Hierarchy Process. Qiang He, Jun Han, Yun Yang and John Grundy Hai Jin, "QoS-Driven Service Selection for Multi-Tenant SaaS", published and proposed a criterion for selecting a SaaS based on QoS parameters like Cost, Response Time, Availability and Throughput. This method for service, selection was multi tenant oriented.Pang Xiong Wen, Li Dong ,"Quality Model for Evaluating SaaS Service",published proposed, an advanced quality model measures the security, software quality of the SaaS and quality of service from the perspective view of a service-provider and service-user independently. Based on this proposed an evaluating model which classify the SaaS service into four levels, including basic level, standard level, optimized level and integrated level. By using the quality model and evaluating model, the customer can evaluate the provider and the provider can use it for quality management. Again, this model was based on Key Feature of SaaS.Xianrong Zheng, "CLOUD QUAL: A Quality Model for Cloud Services", published in the proceedings of IEEE TRANSACTIONS ON INDUSTRIAL INFORMATICS. In this paper, the Author has been considered a service perspective and initiates a quality model called as *CLOUDQUAL* for all type of cloud services. in this model authors proposed quality model with set of dimensions and metrics that are applicable to all services on the cloud. *CLOUDQUAL* contains six set of dimensions for quality are security, elasticity,usability, availability, responsiveness and reliability. Out of all six dimensions usability is considered as subjective and all others have been considered as a objective. The author demonstrated with the help of Azure Blob, Aliyun OSS and Amazon S3, cloud storage services.
The above described research work have been proposed for SaaS selection and SaaS evaluations are based on different attributes, but many authors have been contributed based on Key features and quality attributes in order to measure the quality of SaaS on the Cloud Computing. Even though there are some limitations and restrictions and challenges have been found. In order to to face few challenges the new model is proposed for SaaS Evaluation inspired by the SERVQUAL and CLOUDQUAL for measuring general cloud services. The Proposed model in this paper is focusing only on the software services on the cloud computing environment and it is termed as SAASQUAL (Software as a Service Quality model).

## III. Address Challenges And Concerns of SaaS

The new quality model is proposed to evaluate SaaS(Software as a Service) on the Cloud. The cloud computing software service is completely different from conventional software services. For conventional software services the standard mode is given to measure the internal characteristics and characteristics of Software quality that is called as ISO/IEC 9126 Quality model. This standard was more help full for conventional software service providers in order to measure their software product quality to full fill service user requirements and also service user to identify the potential service as per their requirements. The ISO/IEC 9126 Quality model is explained in the following section.

### 1).ISO/IEC 9126 Quality Model for Conventional Software Services

The standard quality model intended for software products is ISO/IEC 9126 standard. This standard describes the software quality in the form of characteristics. the first part of standard ISO/IEC 9126 presented a classification of software quality in a form of structured set of characteristics and sub-characteristics as shown in the below figure 1. .All quality sub-characteristic is classified into several attributes. An attribute is an property that can be certified or evaluated in the software product.

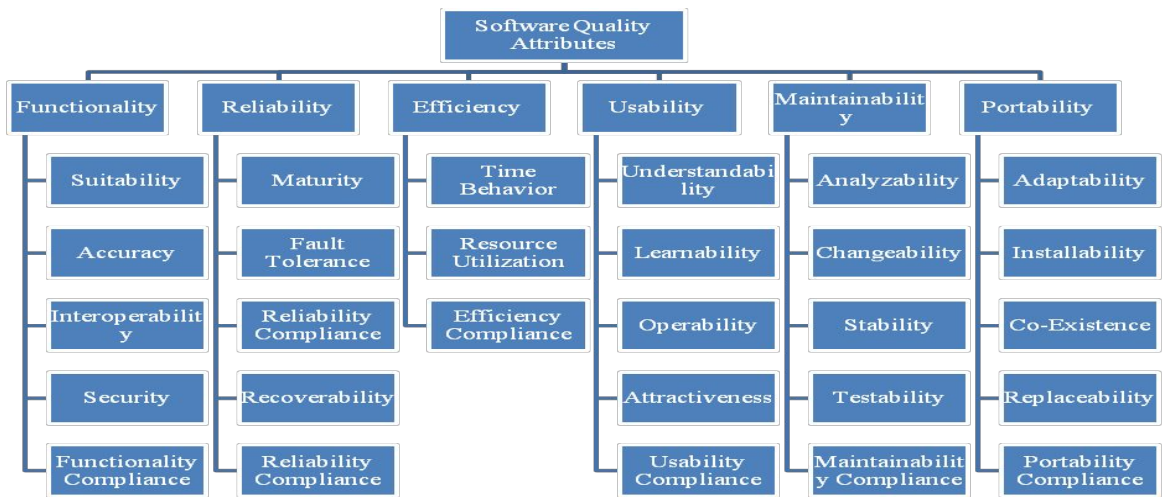

Figure 1: ISO/IEC 9126 Conventional Software Quality Model

### 2)SaaS Development Life-cycle (SaaS DLC)

The service development on cloud computing be in need of a different approach than conventional software development life-cycle,because SaaS development life-cycle becomes a significant success feature of the whole project for cloud providers. Within conventional software development surroundings, further prominence is to settle on the functional characteristic. Since it is situated on an on-premise infrastructure with strong inherent security, dominance, operational simplicity,conformity and alleged service magnitude requirements. An additional significant issue is the cost of operations. It has been repeatedly taking back seat , particularly at cost centers, due to the drop cost and straight payment models. The foremost objective of SaaS-DLC section is intended to achieve center of attention on the life-cycle features of SaaS service development. this sections is also describes the inputs ,motivation and deliverable of all phases of life-cycle. Cloud services have built for inside utilization as well as for disposing to outside customers.In order to develop cloud services for outside consumption, as these development life-cycle needs meticulous architecture implementations to incorporate the service creed mandatory for a victorious business model for services. therefore,the SaaS (Software as a Service) Development Life Cycle demonstrated here is to asses scope at out-most facing services.the process can simply adopted to inside and private cloud based applications that aims at internal users. It has been suggested that every IT enterprise have to begin glancing at themselves as service providers and take step correspondingly. The SaaS Development life-cycle is different from traditional SDLC as depicted in figure 2.

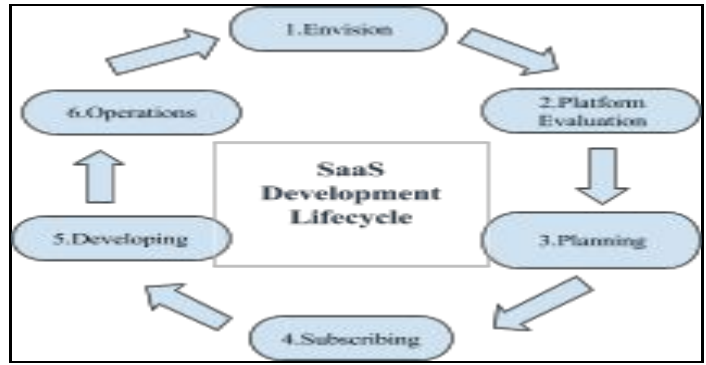

Figure 2: SaaS Development Life-Cycle (SAAS- DLC)

### 3) CLOUD SERVICE PRINCIPLES

In order to use cloud services effectively to meet the requirement goals and specific targets in economically feasible manner,it is necessary for cloud services to be implemented on strong base.the strong foundation that exhibits definite principles of cloud services.The principles have been identified for cloud services are different than conventional services and described as discoverable, detachable, economics, scalability, and supportability.about all these mentioned principles are discussed in the below section.

- ✓ *Discoverability:* It is a service that can be taken care by service users with less human interaction on the part of the service provider throughout complete service lifecycle.this principle gives direction to service users by minimizing the gap between deployment and its vision.The major advantage of service discoverability is that it is minimizing the price of deal considerable expected to network outcome generated through the automation engines assist by service architecture.
- ✓ *Reachability:* This principle is available for all and helps to service users bridge bigger enterprises and small business similar way.on the Cloud platform choosing service provider and suitable service architecture will entertain significant role in activating reachability of SaaS.
- ✓ *Economic Feasibility:* This principle will help to measure consumer usage of service on the cloud at different scales.any cloud service has to be affordable to operate and use.The service subscription amount should be higher than the grand total of the cloud service utilization cost both direct and indirect.Cloud service providers required to assist SaaS architects with cost-oriented service architecture (CSA) methods that include economical assets utilization as the fundamental principle.
- ✓ *Scalability:* this principle demonstrates that cloud services have to supply the multi-tenant platform,which component needed to convey constant performance throughout various conditions of use. In a cloud deployment all virtualized storage components not supporting scalability constraints. the same considerations are verifiable for network and compute resources.
- ✓ *Supportability:* This principle helps frequently to take a back respective to functionality in applications. The cloud service architecture should take care of suitability as one of underlying creed that gives direction towards designing solution and implementation. Incorporation to application subjects, supportability of SaaS should provide peak priority to availability, recovery, performance and disaster recovery because of the outward hosting environment.

## IV. SAASQUAL: Proposed Quality Model

The cloud computing providing many services.every cloud service should possess some features.The general features of cloud services is shown in the below figure 3.

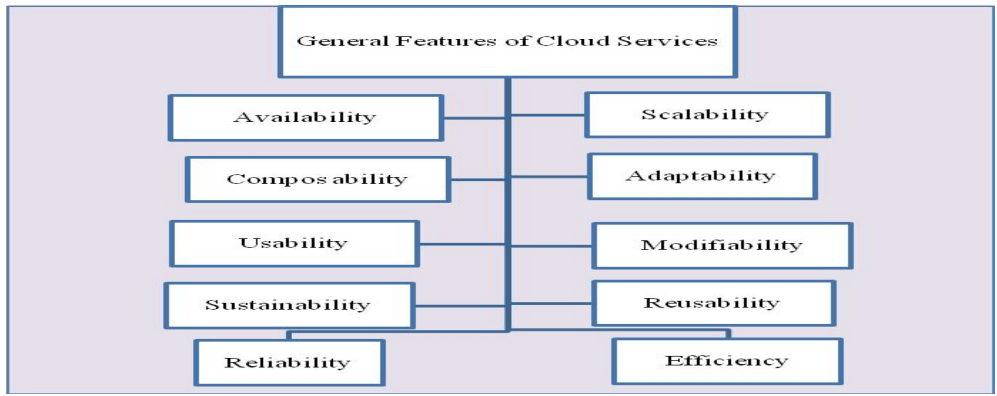

Figure 3: General Features of Cloud Services

On the cloud computing environment there are some key features mostly associated with software services. The essential trademark of SaaS is to recognize a evaluating quality model for SaaS as shown in the below figure5 and it has become a essential to point SaaS essential features. From several meticulous evaluation quality methods proposed by many researchers.[1],[2],[7],[9]. the SaaS Key features are recognized as *Reusability, Availability, Scalability , Paya for use, Customizability, Data Managed by Providers.*

### 1)Model Based Clustering Method

The proposed model is using clustering technique consists of forming groups of objects that possess similar characteristics. clustering is the method describes about similar objects in the Intra-cluster and dissimilar objects in the inter-clusters. Attention towards clustering has been increased due to its many applications in various knowledge domains.There are so many clustering algorithms in data mining,out them the *Model based Clustering* algorithm hypothesize a model intended for every object belongs the clusters also identifies the foremost suitable of the data. A model based clustering technique will discover clustering using a density function that consider the data items sparisity in each cluster.It also gives direction to find number of clusters depending on standard statistics automatically.one of the significant method under model based is EM clustering algoritham.EM clustering works based on the principle of maximum likelihood of unnoticed variables.

### 2)The EM Clustering

The EM algorithm is a type unsupervised learning clustering algorithm, that doesn't need any training data stage.This algorithm is working rely upon mixture models of data mining clustering algorithms. It has been following an iterative process and sub-optimal method to identify the parameters within the probability distribution that has highest chance for its elements.The steps of EM clustering are

- Given data set 'x' is input to algorithm.
- The complete number of clusters represented by 'M'
- 'e' is the acceptable error for the maximum number of iterations.

For every iteration, initially, the E-Step (E-expectation) is executed, that gives estimation on the probability of each and every point suited to which cluster.Secondly the M-step (Maximization) is executed, that helps to re-estimates the parameter vector of the probability distribution of every cluster. Finally the EM clustering method stops when the distribution converge is met or stretch out the highest number of specified rotations or iterations.

The Proposed quality model used to evaluate Software as a service on the cloud. Initially the data related to Software service should be collected from various SaaS users with respect their feedback about to product quality and stored in the form of relations. Then separate the SaaS key features data give on the scale of 1 to 10 rating so that the details available as a numeric data. Next, apply the EM model based clustering algorithm on SaaS data and form a cluster. The clusters will form based on key features of SaaS. Each cluster quality can be measured by using some standard methods for quality measuring of Clustering. The Quality Cluster will help to decide cloud user to select the most desirable features of SaaS and SaaS providers can improve their Product as per the clustering results suggestion to achieve enhancement of their product in order to retain customers in the competitive world. The workflow of SAASQUAL is shown in the below figure 4.

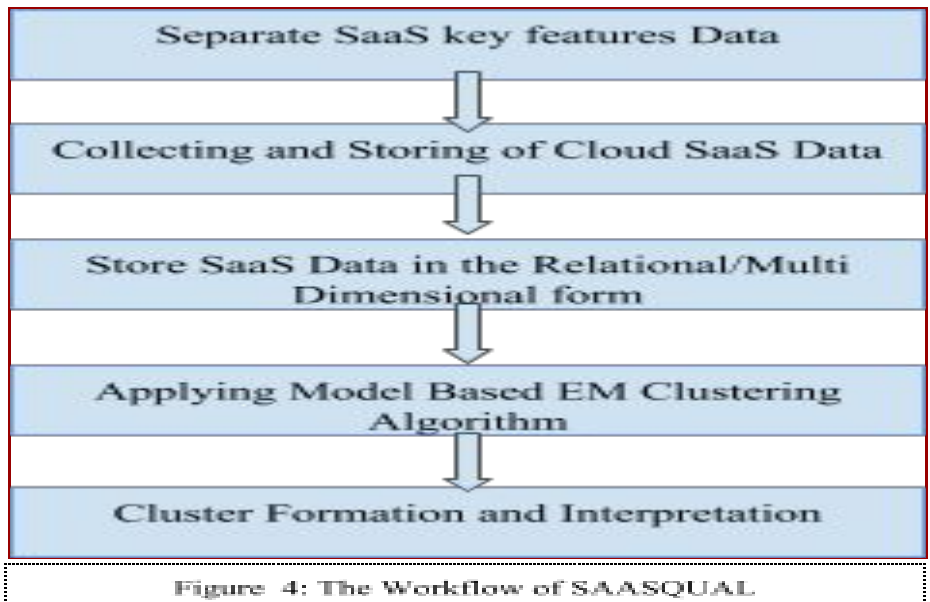

Figure 4: The Workflow of SAASQUAL

## V. Conclusion and Further Enhancement

This paper described about a new quality model for evaluating SaaS on the cloud computing environment. The SAASQUAL model is useful for service users as well as service providers in the cloud to select SaaS and to provide SaaS as per user requirements. Further, it is proposed to use R tool for EM clustering and showing different clusters based on the key features of SaaS[1] and associated metric to measure each quality attributes.In the future i has been intended that the proposed model can be implemented as automated tool for evaluating SaaS Quality.